\documentclass[preprint2]{aastex}
\slugcomment{Astrophysical Journal, October 10 issue}
\lefthead{El-Zant, Shlosman \& Hoffman}
\righthead{Dark Halos: the flattening of density cusp}  
\input epsfig

\def \m3{{\rm Mark III}}
\def \etal {{\it et al.\ }}

\def \ie {{\it i.e.\ } }
\def \eg{{\it e.g.\ }}

\def\bV{{\bf V}}

\begin{document}

\title{DARK HALOS:\\ THE FLATTENING OF THE DENSITY CUSP BY DYNAMICAL FRICTION}

\author{Amr  El-Zant and Isaac Shlosman}
\affil{Department of Physics \& Astronomy, University of Kentucky, Lexington,
   KY 40506-0055, USA \\ email: {\tt elzant@pa.uky.edu} and {\tt
   shlosman@pa.uky.edu}}
\and

\author{Yehuda Hoffman}
\affil{Racah Institute of Physics, Hebrew University, Jerusalem, Israel
       \\ email: {\tt hoffman@vms.huji.ac.il}}

\begin{abstract}
N-body simulations and analytical calculations of the gravitational
collapse in an expanding universe  predict that halos should form with
a diverging inner density profile, the cusp. There are some observational
indications that the dark matter distribution in galaxies might be
characterized by a finite core. This `core catastrophe' has prompted a search
for alternatives to the CDM cosmogony. It is shown here that the discrepancy
between theory and observations can be very naturally resolved within the
standard CDM model, provided that gas is not initially smoothly distributed in
the dark matter halo, but rather is concentrated in clumps of mass 
$\geq 0.01 \%$ the total mass of the system. 
Dynamical friction acting on these lumps moving in
the background of the dark matter particles, dissipates the clumps orbital
energy and deposits it in the dark matter. 
Using Monte-Carlo simulations, it
is shown that the dynamical friction provides a strong enough drag, and that
with realistic baryonic mass fractions,
the available orbital energy of the clumps is sufficient to heat the halo and
turn the primordial cusp into a finite, non-diverging core --- overcoming
the competing effect of adiabatic contraction due to gravitational influence
of the shrinking baryonic component. Depending on the initial conditions,
the total density distribution may either become more or less centrally 
concentrated. Possible consequences of the  proposed mechanism for other 
problems in the CDM model and for
   the formation and early evolution of the baryonic component  of
galaxies are also briefly discussed.  
\end{abstract}

\keywords{galaxies: evolution -- galaxies: ISM -- galaxies: kinematics \&
dynamics -- galaxies: structure -- hydrodynamics}

\section{Introduction and motivation}

The formation and structure of dark matter (DM) halos around galaxies
and clusters of galaxies poses one of the great challenges to theories
of structure formation in an expanding universe. The basic theoretical 
paradigm was set by Gunn \& Gott (1972) and Gunn (1977), and a more detailed
analysis considering  the dependence on the primordial
perturbation field was done by Hoffman \& Shaham (1985). The first  
N-body simulations designed to probe the structure of dark halos indeed 
confirmed the basic findings of Hoffman and Shaham (Quinn, Salmon 
\& Zurek 1986; Frenk \etal\ 1988), but could not resolve
the innermost structure of the dark halos. Higher-resolution simulations 
showed evidence of a cuspy density profile of the form $\rho(r) \propto
r^{-\alpha}$ with $\alpha \geq 1$, down to the resolution limit of the
simulations (Dubinski \& Carlberg  1991; Warren \etal\ 1992; Crone \etal\
1994), and further limited the exponent of the innermost density
profile to $\alpha \approx (1 - 1.5)$ (Klypin \etal\ 2000; Navarro, Frenk
\& White 1997, hereafter NFW; Cole \& Lacey 1996; Tormen, Bouchet \& White
1997; Huss, Jain \& Steinmetz 1999; Fukushige \& Makino 1997; Moore \etal\
1999; Jing \& Suto 2000). NFW suggested a particular parametric density
profile, claimed to be a universal one (independent of halo mass), 
and characterized by an $\alpha=1$
cusp and $r^{-3}$-like decay at large radii.  
On the theoretical side it has been demonstrated that a
cuspy density profile arises inherently from the cold gravitational
collapse in an expanding universe (Lokas \& Hoffman 2000).

Whether the predicted density cusp is supported or ruled out by observations
is subject to debate. Moore (1994) and Flores \& Primack (1994) argued that
the observed HI rotation curves of gas-rich dwarf galaxies favor a
core-halo structure rather than a diverging inner density profile. This 
has been confirmed by more recent observations of De Blok \etal\ (2001).
The DM distribution inferred for spiral galaxies also show large constant
density cores (e.g., Salucci \& Burkert 2000). Moore \etal\ (1999) coined this
disagreement as the `core catastrophe,' using it as a major argument against 
the cold dark matter (CDM) model. Indeed,
alternative theories for structure formation have been proposed, \eg\ 
self-interacting dark matter (SIDM: see,  e.g.,Spergel \& Steinhardt 2000). 
Nevertheless, it
has been suggested that the inconsistency may reflect the finite resolution of
the observations that has not been properly accounted for in the analysis of
the HI rotation curves  (van den Bosch \etal\ 2000; van den Bosch \& Swaters
2000). Moving on to clusters of galaxies, the cuspy density profile agrees with
the recent determination of the mass profile from ASCA X-ray observation of
A2199 and A496 (Markevitch \etal\ 1999). These cannot be accounted for in the
SIDM scenario (Yoshida \etal 2000). In addition, gravitational systems are 
ultimately unstable to heat transport and SIDM models will eventually undergo
gravothermal catastrophe, resulting in even more centrally concentrated structures: 
(Moore \etal 2000; Kochanek \& White 2000).

Assuming that the theory of cold collapse in an expanding universe is
correct, that halos form with a density cusp, and that observations of
galaxies indicate that the present day structure has a finite non-diverging
core, we discuss a mechanism for resolving the above conflict.
Close inspection of the problem at hand reveals that 
adding a mechanism that
enables `heat' transport to the inner part of the halo would resolve
the problem --- destroying the inner cold region where the ``temperature''
inversion occurs.
 This will lead to the puffing up
of the central regions and to the  flattening of the density cusp.  The aim of this
paper is to show that such a mechanism can be naturally formulated within the
CDM cosmogony  without appealing to `new physics.'In addition, as opposed to 
SIDM we have a two component system --- with one component (the halo) that 
is {\em always} expanding and  the other (baryonic) component that is always
shrinking. The baryonic component becoming centrally concentrated is not an 
embarrassment, 
since galaxies do have disk-bulge-central mass components that are
more centrally concentrated than the observed halo structures. In addition, part of
the baryonic mass can be lost in subsequent outflow (which is efficient 
in precisely those galaxies without highly concentrated baryonic components).

The basic scenario envisaged here is one where the gas (\ie baryon)
distribution becomes lumpy because of gravitational collapse and cooling on
its Jeans scale.  Assuming these lumps of gas to be compact enough and their
number densities small enough, so that tidal disruption and collisions are
avoided over at least a  few crossing times, one can consider them as
super-particles moving through the smooth (say NFW) background of DM
particles. Super-particles  experience   dynamical friction (DF), losing
energy to the inner dark halo and heating it up. The DF dissipates the energy
of the gaseous component, but unlike in the case of  radiative dissipation, the energy is
conserved.  Energy lost by the gas through DF is deposited in the inner dark
halo and heats it up. For the above
scenario to work, we have to establish that: 
{\it (a)} the coupling via DF is
strong enough to produce,
within a few dynamical times,
significant effects 
and {\it (b)} with realistic gas mass fraction, there  is enough
energy to heat up the inner halo and to form a core, overcoming the competing
effect of the deepening potential well by the shrinking gas distribution.
The semi-analytical Monte Carlo model used is described in
\S\ref{sec:model}
and the results are given in
\S\ref{sec:results}. 
A general discussion of the implications and ramifications
of the proposed model is found in
\S\ref{sec:discussion}.

\section{Method and Model Parameters}
\label{sec:model}

A rigorous modeling of the dynamics of collapse and virialization of
the DM-gas system demands hyd-rodynamical/N-body simulations starting
from cosmological initial conditions and  with a
very large dynamical range, so that the DF is  properly modeled and the gas
dynamics allows for the fragmentation of the gas component into individual
massive clumps. Present day capabilities, however, only allow for a more
modest approach. One could, for example, perform high-resolution simulations
(e.g., Moore \etal\ 1999; Klypin \etal\ 2000), 
and introduce {\it ab initio} a population of massive super-particles
moving through the background DM. Instead, we model the proposed processes in
a semi-analytical Monte-Carlo approach, in what amounts to  a feasibility
study of  the proposed mechanism.

Consider a  mass $M$ moving with a velocity $\bV_M$ through a homogeneous
system of non-interacting particles of mass $m \ll M$ and density $\rho$
with a Maxwellian velocity distribution of dispersion $\sigma$. The massive
particle experience the DF deceleration  (e.g., Chandrasekhar 1943; Binney and
Tremaine 1987) 
%
$$\left(\frac{d \bV_M}{dt}\right)_{\rm DF}=$$
\begin{equation}
- \frac{4 \pi G^{2} \ln \Lambda
    \rho M}{V{^3_M} } \left( {\rm erf}(X) - \frac{2 X}{\sqrt{ \pi}}
    \exp\left[- X^{2}\right] \right) \bV_M,
\label{eq:df}
\end{equation}
where $X\equiv \bV_M/\sqrt{2}\sigma$, the Coulomb logarithm $\ln\Lambda$ is
defined by $\Lambda=b V{^2_M} / G (M+m)$ and $b$ is the maximal impact
parameter of the two body encounters, usually taken to be the radius of the
system. Although highly simplified and mathematically incomplete, this purely 
local treatment of DF seems, in many situations of interest,
to give results  which are
a reasonable approximation to the true behavior 
(e.g. Zaritsky \& White 1988).
The rate of energy loss by a super-particle is $\dot{E}= M({d  
\bV_M}/dt)\cdot \bV_M$.
This defines a timescale, typically significantly larger
than the dynamical timescale and related to it through the 
ratio of total mass to $\eta =M_{tot}/ M$. More precisely,  
at any given time $X\approx 1$,
allowing one to approximate it by 
$\tau_{\rm DF}/\tau_{\rm dyn} \approx (0.75/{\ln
\Lambda}) \eta$, where $\tau_{\rm dyn}$ is the dynamical 
timescale.  In  virial equilibrium
$\Lambda \approx \eta$, hence $\tau_{\rm DF}/\tau_{\rm dyn} \approx
0.75\eta/{\ln\eta}$. The local DF timescale will differ significantly 
from this and will be much smaller in the inner dense regions.

The mathematical model used here is that of a spherical halo of  
mass $M_{DM}$ with a  population of $n_g$ super-particles of mass
$M$. While CDM halos found in N-body simulations are not perfectly
spherical, they are definitely not flat objects and can,
for our purposes, be reasonably modeled using spherical symmetry. 
Moreover,
the  gaseous component will be assumed to be in the form of compact clumps
with sufficiently small crossections and extended spatial distribution so
that, for the timescales  considered,  dissipative collisions leading to the
flattening of the gas to a preferred plane can be ignored.
In  spherical symmetry, each particle $i$ represents a smoothed shell  
of the same mass $m_i$ and at the same radius $r_i$. Thus, upon sorting the
particles according to their distance from the center, the force on the
$n^{th}$ particle is given by $\Sigma_{i=1}^{n-1} Gm_i/r{^2_n}$. To avoid 
numerical instability we use a Plummer softening of 0.1 \% the initial halo
size.

The numerical model employed here is that of $N_{DM}$ and $N_g$  particles
representing the spatial distributions of the DM and gas 
(\ie super-particles), respectively. We choose to determine $N_{DM}$ and $N_g$ 
solely by considerations  of a reasonable statistical representation of a
two-fluid DM and gas system, coupled via DF. Thus all particles have the same
mass $m=1/N$, where $N= N_{DM}+N_g$. In the limit of $M\gg m$, the dynamical
interaction is independent of the value of $N_{DM}$. To complete the
statistical treatment, a multiplicative factor $F$, the `coupling constant,'
is introduced in Eq.~(1). It determines $M$ and $n_g$ in a system of total
mass $M_{tot}$ and gas mass fraction $f_g$ with the same DF interaction
strength as our corresponding  statistical representation in terms of $m$ and
$N_g$. The correspondence is made through the following relations: $f_g=N_g/N$,
$M =F m M_{tot}$ and $n_g = N_g / F$. For all the models described  here the
particle numbers are  fixed at $N=10^5$ and $N_g= 10^4$. Therefore, $f_g=0.1$.
The code is written in units of $GM_{tot}=1$, where $G$ is the gravitational
constant, so that $m=10^{-5}$. The initial radius of the halo is taken to be
$100$ in these units.

The initial DM distribution is constructed to follow the NFW density profile 
$\rho=\rho_s r{^3_s}/r(r_s+r)^2$ with $r_s$. Once the total mass is fixed,
determining $r_s$ or $\rho_s$ completely determines the halo model. We choose
to fix the former quantity (values are given in Table 1). The scaling to
physical units exploits the ``universal'' scaling relations of NFW and is
given by 
\begin{equation}
[r]=1.63 M{_{DM}^{1/3}}  h^{-2/3} 10^{-4}\ {\rm kpc}
\end{equation}
and 
\begin{equation}
[t]= 0.97
h^{-1} \sqrt{M_{DM}/M_{tot}}\ {\rm Myr}, 
\end{equation}
where $M_{DM}$ is in solar mass and
Hubble constant is defined by $H_0 = 100 h\ {\rm km\ s^{-1} Mpc^{-1}}$. Since
NFW halo density profiles have  the same functional form irrespective of mass,
the mass unit can be  arbitrarily chosen. The system of units are completely
fixed once the gas mass  fraction is. One hundred time units will correspond
to about 11 central  (in the region $r < 10$) dynamical times and the length
unit will be about  1~kpc for a halo of $5 \times 10^{11} {\rm M_\odot}$.

\begin{deluxetable}{lcccccc}
\tablecaption{Model Parameters}
\tablehead{
Model &  $F$ & $r_s $ &
       Gas at $t=0$ & $n_g$ & $M/M_\odot$ (for $M_{DM}=10{^{12}M_\odot}$)
}
\startdata
{\bf 1}  & $20$              & $5$  & uniform $R_g=20$  & 500 &
  $2\times 10^8$                       \nl
{\bf 2}  & $100 \times e^{-T/400}$ & $5$ & uniform $R_g=20$ & $100\times
e^{T/400}$  &   $10^9 \times e^{-T/400}$ \nl
{\bf 3}  & $ 20        $     & $5$  & NFW  & 500
&   $2 \times 10^{8}$            \nl
{\bf 4}  & $ 100       $     & $3$  & NFW  & 100
&   $10^{9}$                     \nl
\enddata
\label{table:models}
\end{deluxetable}          

Each Cartesian velocity component is sampled from a Gaussian distribution
with zero mean and variance $\sigma =\sigma(r)$, where $\sigma$ is obtained by
solving the steady state Jeans equation for the radial velocity dispersion,
the solution of which can be obtained in terms of elementary functions and
quadrature (with boundary condition $\sigma=0$ as $r \rightarrow \infty$). Gas
particles are either sampled from the same distribution or are taken to be
homogeneously distributed within a certain radius $R_g$. In the latter case,
their initial velocities are assumed to have a  Maxwellian with half of the 
halo dispersion at $R_g$, and  the combined system is then scaled to virial
equilibrium. Systems are left to relax for 600 to 1000 time units 
(corresponding to about  6 to 10 central dynamical times) without including 
the effect of the DF.

The density and velocity dispersion of the DM particles are calculated in a
1000 bins of equal numbers and these values are used in evaluating the DF
acting on the $N_g$ gas particles using Eq.~(1). The  average energy lost by
the gas particles in a given bin per unit time is updated at fixed time steps
$\Delta t=5$.  Multiplied by $\Delta t$, this is the energy to be gained by
the DM particles of the same bins. At the end of each time interval, the
Cartesian velocity components of the DM particles are updated through an
additive term chosen from a normal  distribution with zero mean and variance
$\sqrt{(2/3) E_{b}/m}$. Here $E_{b}$ is the energy gained per DM particle.

The integration is advanced using a Runge-Kutta-Merson method with adaptive
timestep and predetermined local tolerance. The results were found to
converge for  tolerances $\leq 10^{-3}$.  Because of the strict spherical
symmetry, the gravitational field is much smoother than in the 3D case and 
evolutionary effects are also very slow. After reaching dynamical equilibrium
(and in the absence of frictional forces), a system of a few thousand
particles may remain virtually unchanged for hundreds of dynamical times. The
total energy is conserved to five digits in the absence of DF. When the DF
is turned on, the  statistical nature of the energy feedback and finite size
of the shells causes additional fluctuations. Still the change in the energy
did not exceed $0.51 \%$ of the total energy, or a few percent of the energy
{\em exchanged} through the DF, and was not systematic.

The results are shown for times up to $T = 1800$ (where $T=0$ is taken to be
the  time when the DF starts acting) corresponding to $\sim 2$~Gyr (or
about 18 central dynamical times), although the models have been advanced
much further.  We anticipate that eventually the clumps' evolution will become
dominated by collisions, leading most probably to their destruction through
star formation, merging, fragmentation, and other processes which will provide
a natural termination to the DF action. For example, for $n_g$  objects of 
radius $d$ randomly moving inside a radius $R_G$, the mean free time is $1/3
n_g (\frac{R_g}{d})^{2}$ crossing times. For the parameters of Model~1 (c.f. 
Table~\label{table:models}) and assuming   $d\sim r_s/50$ for example,  this 
is about 25 crossing times. Detailed analysis of the aforementioned processes
is outside  the scope of this paper. Here, we either terminate the evolution
at $T=1800$ abruptly (Models~1, 3, 4), or introduce an exponential cutoff 
(with characteristic timescale of 400 units) in the coupling parameter 
$F$ (Model~2).

\section{Results}
\label{sec:results}

A wide range of models have been calculated but only four models are
presented here, the parameters of which are given in Table \ref{table:models}.
These parameters are representative yet not exhaustive of the types of
systems we have studied. The general trend is the same for all the models:
the DF is found to be an efficient mechanism of heating up the DM halo,
causing the flattening of the density cusp into a core-halo structure and
consequently resulting in less steeply rising rotation curves. In principle, 
it is possible to have a much smaller
$N_g \sim 1000$ and to obtain similar results for smaller $R_g \sim 5$.

\begin{figure*}[ht!!!!!!!!!!!] 
\epsfig{file=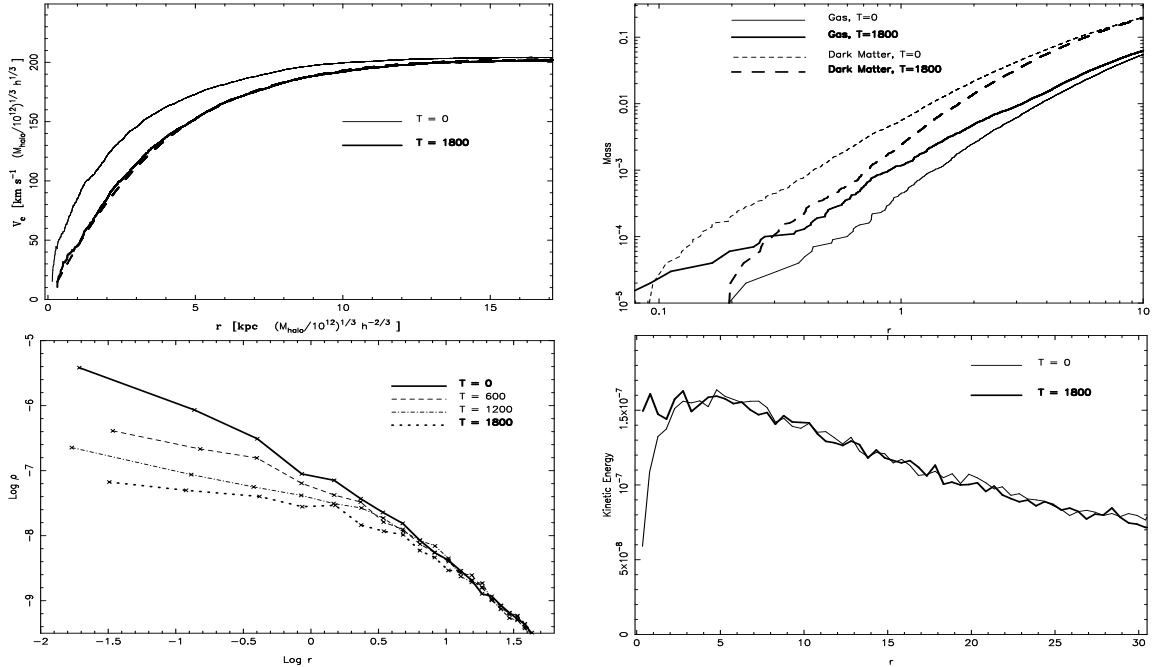,width=17.cm,height=14.5cm,angle=+00}
\caption{Evolution in central region of Model~1. Top left: Initial and final DM
rotation curves. The dashed line is a least squares fit using Eq.~(\ref{burk}).
Top right: Initial and final gas and DM cumulative masses. Bottom left: DM
density at four different times. Bottom right: Initial and final kinetic
energy per DM particle.} \label{fig:mod1}
\end{figure*}

Fig.~\ref{fig:mod1} shows the evolution of quantities characteristic of the 
matter distribution of Model~1.
We note that the final DM rotation curve rises less
steeply  as the DF puffs up the inner halo. To get a  quantitative assessment
of the disappearance of the cusp and the emergence of a core-halo structure,
the final DM rotation curve has been fitted by Burkert's (1995) profile,
\begin{equation}
\rho=\frac{\rho_0 r_0^{3}}{(r+r_0)(r^2+r_0^{2})},
\label{burk}
\end{equation}
claimed to be a good fit to the DM rotation curves inferred from
observations (e.g., Salucci \& Burkert 2000).

Assuming the density to be 
constant, $\rho\approx \rho_0$ inside $r_0$ and using Eq. 5 of 
Burkert \& Salucci, one finds the observational
correlation 
 $\rho_0\propto r_0^{-2/3}$ between the core radius and density. 
We now note  that the change in the rotation curve from a NFW profile to a 
Burkert profile is the result of a decrease in  central density accompanied
by heating of the central region and that the region where this
mechanism is effective lies inside $r_s$ (as can be seen from the lower panels
of Fig.~1). 
Hence, one expects $r_0\approx
r_s$ in the final configuration. In addition, for the density to be fit at
large radii, $\rho_0\approx \rho_s$. Moreover, it can be shown (e.g. by 
using the first four equations of  Burkert \& Silk 1999 and assuming $r_s$ to 
be small compared to the virial radius) that $\rho_s\propto
r_s^{-9/(\gamma+3)}$, where $\gamma =7.14$ for SCDM models, and $\gamma=10$
for CDM$\Lambda$ cosmologies. Therefore, especially for  $\gamma\approx 10$,   
one is  able to anticipate the observational correlation between $r_0$ and
$\rho_0$ from the  theoretical one between $r_s$ and $\rho_s$ obtained from
numerical simulations --- provided that the DF mechanism is able to wash out
the density cusps of halos of all masses. 

\begin{figure*}[ht!!!!!!!!!!]
\epsfig{file=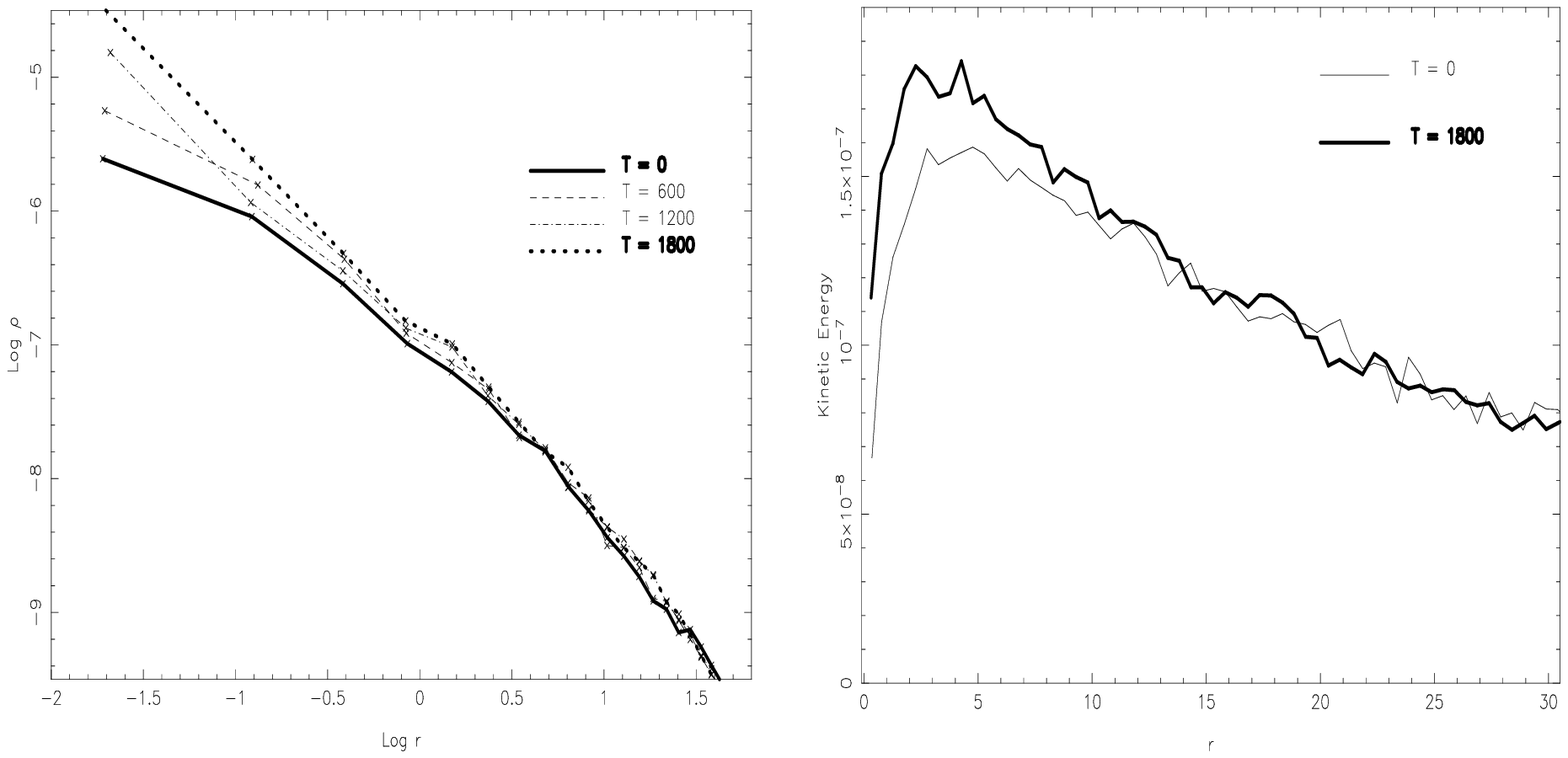,width=17.cm,height=14.5cm,angle=+00}
\caption{Evolution of DM density and kinetic energy per DM particle
in Model~1 in the absence of energy feedback.}
\label{fig:adiabs}
\end{figure*}

For comparison, we have studied systems with no
feedback into the halo from  the gas lumps. 
In this case, as expected the halo contracts
adiabatically, becomes more centrally concentrated 
 and the kinetic energy profile increases self-similarly 
(Fig.~\ref{fig:adiabs}). 
This behavior is thus similar to the case of a smooth gaseous system 
contracting
via dissipational collapse inside the dark matter halo. It is then clear that
the major assumption here, leading to the expansion of the host halo, is that
the gas is distributed in clumps. The resulting energy feedback is able to 
overcome the competing effect of adiabatic contraction. The halo thus expands 
instead of contracting further under 
the gravitational influence of the shrinking baryonic component.

\begin{figure*}[ht!!!!!!!!!!!]
\epsfig{file=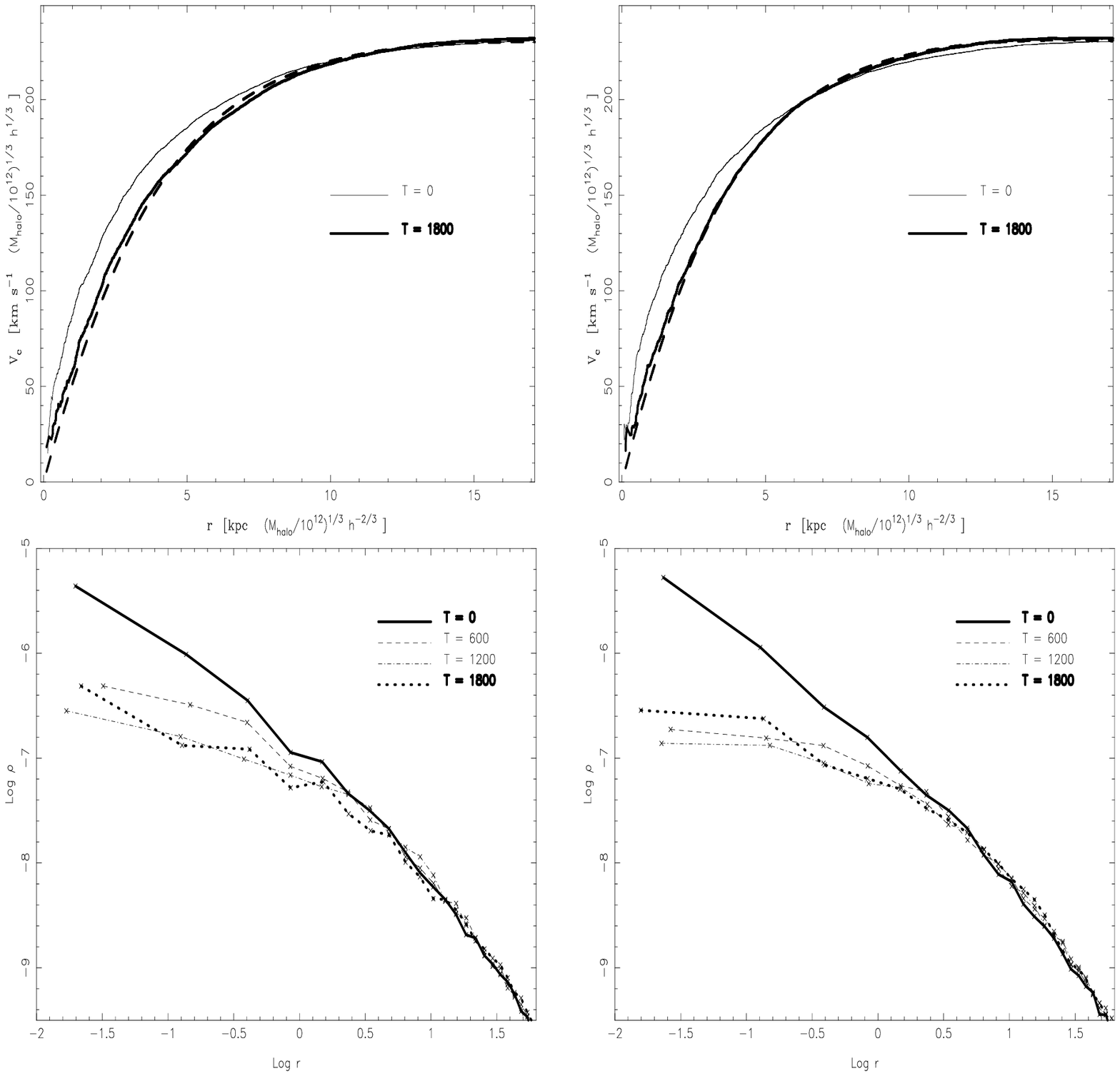,width=17.cm,height=14.5cm,angle=+00}
\caption{ Top: Initial and final total rotation curves of
Model~1 (left panel) and Model~2. Dashed lines represent
fits using Eq.~(\ref{burk}).
Bottom: The total density profile for the same models given at four
different times.}
\label{fig:tots}
\end{figure*}            

It is possible to obtain nearly identical results for larger coupling 
parameters (thus higher clump masses $M$) acting on  much shorter time
scales, as in the case of Model~2, where an exponential cutoff in the
two-fluid coupling (parameter) was implemented. This  can be interpreted
as representing a fragmentation process.In both Models~1 and 2,  the
inflow of clumps due to energy loss via DF is small enough so that the {\em
total} density in the inner regions also decreases and the total rotation
curves can still be fit with Burkert  models, as can be seen from
Fig.~\ref{fig:tots}. 
In Model~1, if our process is allowed to continue, it will ultimately cause
excessive central gas concentration, so that the total density fit to
Eq.~(\ref{burk}) will  worsen. 
This is not true of Model~2 however, 
where due to the exponential cutoff, the
the coupling via the DF is already too small at $T = 1800$
to have any  significant effect. In all cases however, the DM 
distribution continues to be fit well by a core-halo profile far beyond
the timescales shown here.

\begin{figure*}[ht!!!!!!!!!!]
\epsfig{file=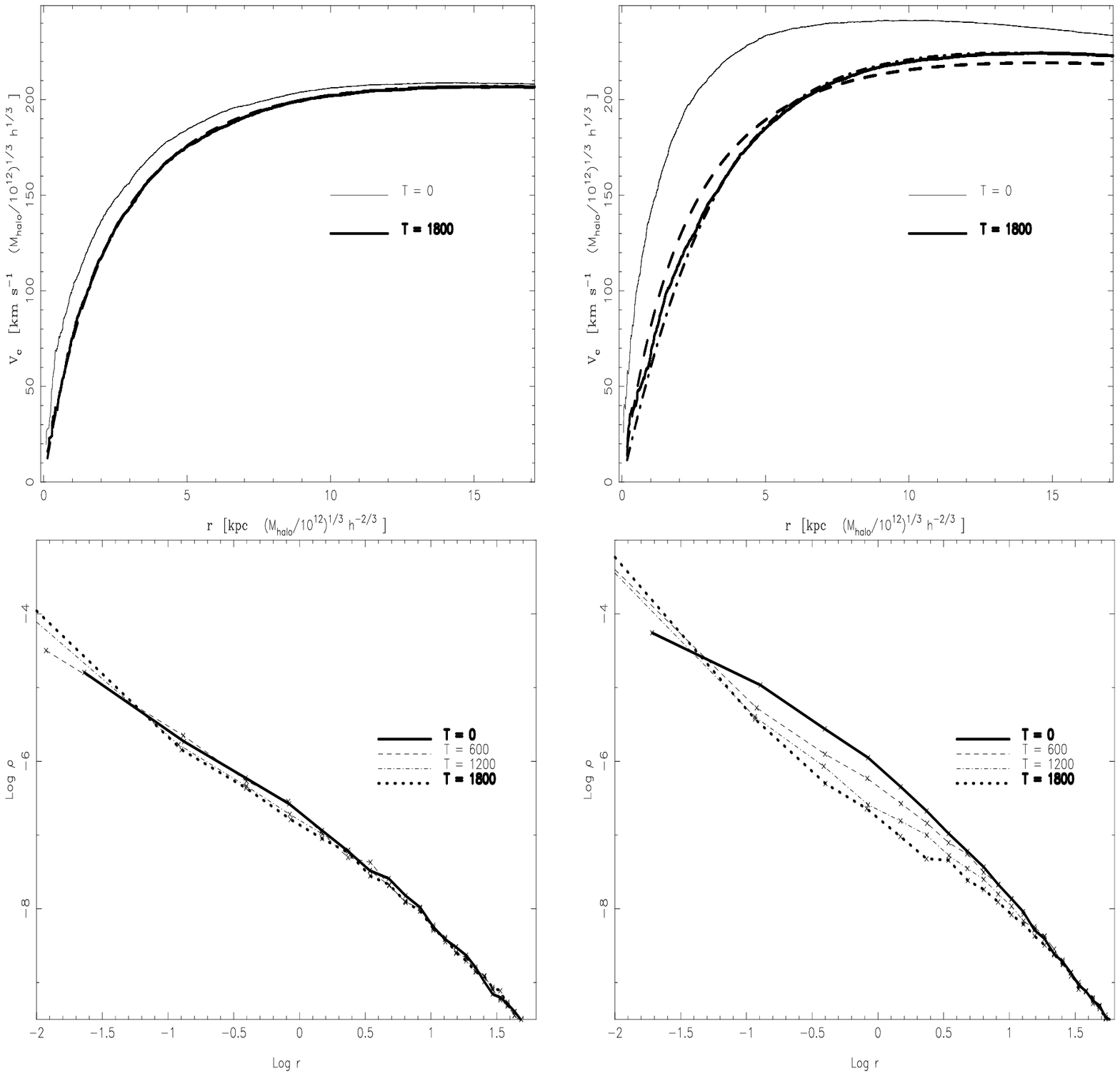,width=17.cm,height=14.5cm,angle=+00}
\caption{Top: Initial and final DM rotation curves for Model~3
(left)
and Model~4. Dashed lines represent best fits using Eq.~(\ref{fit}),
with $R_0=2.12$ and $A=2.14$ for Model~3 and $R_0=2.027$ and $A=2.017$
for Model~4 (in the units defined in section~2).
The dashed dotted line in the right panel represents a fit using
Eq.~(\ref{burk}). Bottom: The total density profiles for the same
models given at four different times.}
\label{fig:mod34}
\end{figure*}

The value of $F$, and hence $M$, controls the rate at which  energy is
transported, but the integral quantity depends mostly on the mass fraction in
clumps in the region of interest (i.e., inside $r \sim 10$). In the above
models, after the initial period of relaxation, the mass in the gas component 
comprises about  a fifth of the mass in this region distributed in a much less
concentrated way than the DM (Fig~\ref{fig:mod1}, upper right panel). 
If the clumps are initially distributed the
same way as the DM, their number density  will be too small to have
a significant effect outside the very central region for $F=20$. 
They also have smaller binding energy to give.
In such cases
it was found that the final rotation curves can be fit by
\begin{equation}
\rho= \frac{C}{(r+r_c) (A+r)^2},
\label{fit}
\end{equation}
which rapidly converges to  the NFW profile for $r>r_c$ --- and the requirement 
that it does so fixes the parameter $C$. Moreover, it was invariably found that
best fits require that $r_c=A$, to very high accuracy, making it a
one parameter fit. To compensate for the smaller gas mass fraction, one can
increase $F$ or the central  density --- which increases the coupling {\em
and} decreases the dynamical time, thus rendering the process much faster.
In such cases, the final rotation curves at $t=1800$ can no longer be well fit by 
Eq.~(\ref{fit}): the DM distribution is now modified at intermediate radii
and is better fit by Eq.~(\ref{burk}) (Fig.~\ref{fig:mod34}, upper panels).
Another consequence of the strong coupling and condensed initial distribution
is that the total density profile is very  centrally concentrated
(Fig~\ref{fig:mod34}, lower panels). This is due to the more concentrated
initial distribution and smaller mass, implying smaller amount of 
binding energy that can be released by the clump system,
within a given radius, before it becomes highly concentrated --- 
it could possibly  lead to the formation of the galactic 
bulge/black hole system.
This and other possible scenarios for the fate of the clumpy gas will be
discussed elsewhere.

\section{Discussion}
\label{sec:discussion}

Starting from the NFW cuspy initial density profile for the halo, we have 
shown that dynamical friction provides an efficient mechanism for 
transporting energy from
the clumpy component to the dark matter and heats it up. Under a wide range of
conditions, the clump orbital energy is sufficient to flatten the density cusp
and to form a core-halo structure. Although we assume that $10\%$ of the total
mass is in the clumpy component, in practice only a fraction of this mass,
initially located within $r \sim r_s$, participates in the process.  This is
sufficient to change the structure of the central, high-density region.  

The general scenario presented here is that of the formation of galaxies by
the collapse and fragmentation of gas in DM halos. Within the CDM paradigm,
the DM collapse is followed by the Jeans scale collapse and radiative cooling
of the gas.  The details of this complicated process are still largely
unknown, but the clumpy nature of the gas is unavoidable in the standard
cosmogony (Hutchings \etal\ 2000). It is also required within the hierarchical
merging scenario of galaxy formation.

For dwarf galaxies, clump masses of $10^{5}$ -- $10^{6}$  ${\rm M_{\odot}}$
are required for the DF to be efficient in destroying the central cusps. 
For large galaxies,  masses of $10^{8}$ -- $10^{9}$ ${\rm M_{\odot}}$
are needed.
Such massive inhomogeneities are expected to appear 
after the formation of the first objects,
when UV radiation and inefficient cooling would prevent the subsequent collapse of
objects with masses smaller than  $10^8\ {\rm M_{\odot}}$ --  $10^9\ {\rm M_{\odot}}$,
as argued by  Haiman, Rees \& Loeb (1997). The basic feature assumed here is that the gas
component is  clumpy and that clumps are bound enough to survive in the DM 
background for  a few dynamical times without colliding or disintegrating.
If indeed, at least some of these massive clumps survive long enough and are
not  tidally stripped or otherwise destroyed, 
they will dissolve the most massive
DM cusps. In this way, more massive halos would end up with the biggest cores --- since
they initially  start with the largest NFW scalelength.
This is in line with what is inferred from observations 
(as shown in  section~\ref{sec:results}).
Alternatively, a mass spectrum of these clumps with a lower mass tail
can, in principle, be created  by a variety of  physical processes. 
The details of
such processes are  outside
the scope of this paper --- although in some of the runs we do allow
 for an exponential cutoff in the gas-DM coupling via DF, thus mimicking
a fragmentation process.

If the mechanism  described in this paper starts acting early in the 
history of the hierarchical merging process, when
the halo masses are small, proportionately smaller clump masses 
would be sufficient in changing the CDM halos' central structures
--- since the efficiency of our mechanism depends on the relative masses
of the clumps and halo component. 
In this case the final core
radii and their variation  with  halo mass will 
depend on the details of the merging process.  
The resultant central structure of these halos,  however, should 
still exhibit 
constant density cores, since these are  conserved by the  merging process
(see, e.g., Pearce, Thomas \&  Couchman 1993).

The model presented here provides a framework for understanding the
possible discrepancy between observations of galaxies and the outcome
of N-body simulations. The final dark halo density distributions are well
fit by empirical profiles inferred from observations. The  less concentrated 
density profiles may also be relevant to the substructure problem --- smaller 
halos with low density cores would be less likely to survive to be observed 
today. For simplicity, and in order to emphasize the feasibility of the 
processes described here --- that the energy deposited in the halo  is
sufficient, given realistic gas mass fraction, in overcoming the effect of
the adiabatic contraction and leading to the formation of a significant 
constant density core --- 
we have considered models with zero net angular momentum. Thus we cannot 
comment in detail as to the effect of our mechanism on the angular momentum
problem present in CDM models of disk formation. In principle dynamical friction 
may also  transfer angular momentum from the baryonic component
to the DM one. Our mechanism however is effective only in the central region 
of galaxies and the clumps are likely to end up  in a relatively low
angular momentum system, such as a galactic bulge-central mass component. 
In the outer regions, 
dynamical friction is not effective. The fate of the gas lumps will be determined 
by collisions. These will lead to collapse to a preferred plane, accompanied by 
contraction and spin up. Once this ``second generation'' material has reached 
the central regions it is likely to loose less angular momentum to the
resulting inner halo --- since the latter is 
less concentrated and  has higher angular momentum 
--- than it would by interaction with the original NFW halo. This could lead 
to less concentrated disks. It is also possible that tidal interactions
between the less concentrated halos lead to higher net angular momenta.

  We find that, under certain circumstances, the
{\em total} rotation curves can also exhibit a halo-core structure, albeit more
modest than the DM alone. As emphasized above, one can also envisage a scenario where the
clumps sink to the bottom of the potential well, dissipating their orbital
energy by the DF and collisions  to form the galactic disk/bulge and the
central black hole. Part or most of the gas can subsequently be blown away by
positive energy feedback provided by massive star formation and supernovae
explosions in the form of galactic winds. The efficiency of this process
should anti-correlate with the depth of the potential well and hence
distinguish between dwarf and massive galaxies.

We use a semi-analytical Monte-Carlo approach, based
on the Chandrasekhar approximation for the DF drag force in a homogeneous
medium of uncorrelated particles. This approximation can be improved by
high-resolution simulations in which the super-particles are introduced  by
construction  and where the full $N$-body interaction, including potentially
important global response of the background medium (e.g., Weinberg 2000) is taken
into account. N-body simulations of the dynamics of satellite galaxies in
massive halos show that the semi-analytical modeling underestimates the
DF effect when the satellite mass fraction is  $0.05-0.2$ 
(Tormen {\em et al.} 1998), similar to the fraction of baryonic mass within
$r<10$ in our models. The eventual slowing down of the self-consistent
evolution compared to  the semi-analytic one in their models is probably due
to the  decreasing central background density resulting from
energy feedback into the background medium from the dissipating 
satellites, which is not modeled in  their semi-analytic calculations.
We have also assumed spherical symmetry. Simulated DM halos are found to
be triaxial. In such systems the effect of DF is strongly enhanced (Pesce, 
Capuzzo-Dolcetta \& Vietri 1992).
Therefore, the effect described in this paper can be probably achieved
for substantially smaller clump masses or on smaller timescales.

\acknowledgments

YH acknowledges fruitful discussions on the validity of the NFW
density profile with Ewa Lokas. AEZ would like to thank Francoise Combes
and Walter Dehnen for helpful discussions and Adi Nusser for commenting 
on the manuscript. 
 This work was supported in part by
NASA grants NAG 5-3841, WKU-522762-98-6 and HST GO-08123.01-97A to IS,
and by by the Israel Science Foundation grant 103/98 to YH.


\begin{thebibliography}{}

\bibitem[]{bur} Burkert, A.,  1995, \apj, 447, L25
\bibitem[]{fwde} Frenk, C.S., White, S.D.M., Davis, M., \& Efstathiou, G.
    1988, \apj, 351, 10
\bibitem[]{cha} Chandrasekhar, S. 1943, \apj, 97, 255 
\bibitem[]{deblok} de Blok, W. J. G., McGaugh, S. S., Bosma, A. \& Rubin, V. C.
2001, preprint (astro-ph/0103102)  
\bibitem[]{fm} Fukushige, T., \& Makino, J. 1997, \apj, 477, L9
\bibitem[]{hrl} Haiman, Z., Rees, M. J., \& Loeb, A. 1997, \apj, 476, 458  
\bibitem[]{hs} Hoffman, Y., \& Shaham, J. 1985, \apj, 297, 16
\bibitem[]{hjs1} Huss, A., Jain, B., \& Steinmetz, M. 1999, \mnras, 308,
    1011
\bibitem[]{js} Jing, Y. P., \& Suto, Y. 2000, \apjl, 529, L69
\bibitem[]{kkbp} Klypin, A., Ktavtsov, A. V., Bullock, J. S., \&
Primack, J. P., 2000, subm. to \apj\ (astro-ph/0006343)
\bibitem[]{kow} Kochanek, C. S. \& White M. 2000, \apj 543, 514
\bibitem[]{lh} Lokas, W.L. and Hoffman, Y., 2000, \apjl,   542, 139
\bibitem[]{MVFC} Markevich, M., Vikhlinin, A., Forman, W. R., and Sarazin,
C. L., 1999, \apj, 527, 545
\bibitem[]{MVFS} Markevitch, M.  1999  astro-ph/9904382
\bibitem[]{mq} Moore, B., Quinn, T., Governato, F., Stadel, J., \& Lake,
    G. 1999, \mnras, 310, 1147
\bibitem[]{mq} Moore, B., Gelato, S., Jenkins, A., Pearce, F. R., Quillis, P.
2000, \apj 535, L21
\bibitem[]{nfw} Navarro, J. F., Frenk, C. S., \& White, S. D. M. 1997,
    \apj, 490, 493 (NFW)
\bibitem[]{pearce} pearce F. R., Thomas, P. A. \& Couchman, H. M. P. 1993,
\mnras, 264, 497
\bibitem[]{pes} Pesce, E., Capuzzo-Dolcetta, R. \& Vietri, M. 1992, \mnras, 
      254, 466
\bibitem[]{qsz} Quinn, P. J., Salmon, J. K., \& Zurek, W. H. 1986,  Nature,
    322, 329
\bibitem[]{tbw} Salucci, P., Burkert, A., 2000, \apj 537, L9
\bibitem[]{tbw} Tormen, G., Bouchet, F. R., \& White, S. D. M. 1997,
    MNRAS, 286, 865
\bibitem[]{tds} Tormen, G., Diaferio, A., Syer D., 1998, \ 
\bibitem[]{brdb} van den Bosch, F. C., Robertson, B. E., Dalcanton, J. J.,
    \& de Blok, W. J. G., 2000,   AJ, 119, 1579
\bibitem[]{bs} van den Bosch, F. C. and Swaters, R. A., 2000, \mnras, 299,728
preprint (astro-ph/0006048)
\bibitem[]{bs} Weinberg , M. D.,  2000, preprint
(astro-ph/0007276)
\bibitem[]{bs} Yoshida N., Springel V., White, S. D. M. \& Tormen, G. 2000,
\apj 544, L87
\bibitem[]{zw} Zaritskty, D., \& White, S. D. M., 1988, \mnras 235, 289

\end{thebibliography}
\end{document}